\documentstyle[sprocl]{article}

\def\be{\begin{equation}}
\def\ee{\end{equation}}
\def\bea{\begin{eqnarray}}
\def\eea{\end{eqnarray}}

\begin{document}

\begin{flushright}
IPM-98-280 \\
hep-th/9803067
\end{flushright}

\title{Mixed Branes and M(atrix) Theory on Noncommutative Torus}

\author{F. Ardalan, H. Arfaei, M.M. Sheikh-Jabbari} 

\address{Institute for studies in theoretical Physics and Mathematics 
IPM\\ P.O.Box 19395-5531, Tehran, Iran\\
E-mail: ardalan, arfaei, jabbari@theory.ipm.ac.ir}


\maketitle\abstracts{ 
We derive the noncommutative torus compactification of M(atrix) theory
directly from the string theory by imposing mixed boundary conditions on the 
membranes. The relation of various dualities in string theory and M(atrix) 
theory compactification on the noncommutative torus are studied.}

\section{Introduction}
 
 Compacitification of the M(atrix) model on tori $T^{d}$ [1] have lead to 
interesting consequences, including super Yang Mills theories on d+1 
dimensional dual tori for lower dimensions and non trivial theories for 
higher dimensions. It was observed recently [2] that compactification of the 
M(atrix) model on a two dimensional "noncommutative"
torus $T^{\theta}$, leads to a non local gauge theory defined on a 
"noncommutative" torus. Here the torus $T^{\theta}$ is "noncommutative" 
in the sense of A. Connes' noncommutative geometry [3], defined by the 
noncommutative $C^{*}$ algebra generated by the elements $U_{1}$,$U_{2}$ with,
\be
U_1\ U_2=e^{2i\pi\theta}\ U_2\ U_1.   
\ee
$\theta$ being the noncommutativity parameter of the torus $T^{\theta}$. 
It was then observed [2,4] that this compacitification on $T^{\theta}$ is
equivalent to an M theory three form C, with nonzero value for one of the 
indices in the minus direction $C_{-12}$.

Genuine noncommutativity of space-time in the string theory and D-branes was 
first observed in [5], where coordinates of  D-brane embedding become 
noncommutative.
In M(atrix) model, however, the large scale coordinates of the center of mass
of the system 
are still commuting variables, and remain so for the ordinary toroidal
compactification.

It is therefore important to understand fully the connection among the various 
appearances of noncommutativity in string theory, M-theory, and M(atrix) model. 
In this paper we will make one such connection, relating the Connes' 
noncommutative geometry's
resurgence in M(atrix) compactification 
[2] to a phenomenon of noncommutativity
of space due to mixed boundary conditions in string theory [6,7].
In [7], it was shown that in the presence of a Kalb-Ramond antisymmetric 
field $F$, 
strings satisfying mixed boundary conditions at $\sigma=0,\pi$,
have noncommuting center of mass coordinates.
It was then observed in [7] that a moving membrane in M-theory, results in 
the above string configuration.
The similarity these two kind of noncommutativities  
cries for an explanation. In this paper we attempt to provide one. 

\section{Mixed Boundary Conditions}

Following [6,7] we consider the action
\be
S= {1 \over 4\pi\alpha'} \int_{\Sigma} d^2\sigma \bigl[\eta_{\mu\nu}
\partial_aX^{\mu}\partial_bX^{\nu}g^{ab}+ \epsilon^{ab} B_{\mu\nu}
\partial_aX^{\mu}\partial_bX^{\nu}\bigr] +
{1 \over 2\pi\alpha'}\oint_{\partial \Sigma} d \tau A_i \partial_{\tau}\zeta^i, 
\ee
Variation  of $X^{\mu}$ gives the boundary conditions for $\sigma=0,\pi$, with
\be
F=F_{12}=B_{12}-A_{[1,2]}.
\ee
We next impose the canonical commutation relations on $X^1,X^2$ and their 
conjugate momenta i.e. $P^1,P^2$: 
$$
[X^{\mu}(\sigma,\tau),P^{\mu}(\sigma',\tau)]=i\eta^{\mu\nu}\delta(\sigma-\sigma').
$$
leading to the nontrivial result
\be
[X^1(\sigma,\tau),X^2(\sigma',\tau)]=2\pi i F \theta(\sigma-\sigma'),
\ee
for the space coordinates of the membrane.
Considering the mode expansions for $X^1$ and $X^2$ consistent with our 
boundary conditions, results in the noncommutativity of the center of mass 
coordinates 
$$
x^1=\int X^1(\sigma,\tau)\ d\sigma,\;\;\;\ ; \;\;\
x^2=\int X^2(\sigma,\tau)\ d\sigma.
$$
\be
[x^1,x^2]=-2\pi iF.
\ee

This noncommutativity of space coordinates of the mixed membrane, which 
formally is a consequence of the appearance of momenta $p_i,i=1,2$, in the  
expression for $x^i$, innocent
as it looks, reflects the zero brane distribution inside the D-membrane
[6,7] and therefore is closely related to the noncommutativity of the D-brane
dynamics [5].

\section{Mixed Brane Wrapping}

In order to recover the noncommutativity of the torus of the M(atrix) model 
compactifications, in the presence of Kalb-Ramond field, we need to consider and
understand wrapping of our mixed branes over a torus.
Thereby we will be able to compute the mass spectrum of the wrapped mixed brane 
and find its symmetries, thus reproducing the spectrum and the symmetries 
of the M(atrix)  
model noncommutative torus compactification.

To do so, it is convenient to view the mixed brane as a  T-dual of a
D-string obliquely  wound on a torus. Thus we first take a D-string 
which winds around a torus with cycles of radii $R_1$ and $R_2$ making an 
angle  $\alpha$ with each other: 
\be
\tau={R_2 \over R_1}e^{i\alpha}=\tau_1 +i \tau_2, \;\;\;\;\;\ 
\rho=iR_1R_2\sin \alpha=i\rho_2.
\ee

The D-string is located at angle $\phi$ with the $R_1$ direction
such that it winds $n_1$ times around $R_1$ and $n_2$ times around $R_2$.
Hence
\be
\cot \phi={n_1 \over n_2 Im\tau}+\cot \alpha.
\ee
Imposing the boundary conditions for the open strings attached to 
the oblique D-string, their mode expansions are
\be
X^i=x^i + p^i \tau+ L^i \sigma + Oscil. \;\;\; , i=1,2. 
\ee
where $p^i$ and $L^i$, in an usual complex notation, are:

\be 
p=\ r {{n_1+n_2\tau} \over {|n_1+n_2\tau|^2}} \sqrt{{\tau_2\over \rho_2}}
\;\;\;\;\;\ ; r\in Z.
\ee
\be
L=\ q {\rho{(n_1+n_2\tau)} \over {|n_1+n_2\tau|^2}} \sqrt{{\tau_2\over \rho_2}}
\;\;\;\;\;\ ; q\in Z.        
\ee
As we see, $p$ is parallel to the D-string and $L$ is normal to it. 
It is interesting to note that the length of $L$ is an integer multiple of
a minimum length. This is the length of a string stretched between two 
consecutive cycles of the wound D-string.

Mass of such an open string is found to be
\be 
M^2=|p+L|^2+{\cal N}= {{\tau_2}\over {|n_1+n_2\tau|^2}}{{|r+q\rho|^2} \over 
\rho_2}+{\cal N}.
\ee

The spectrum is manifestly invariant under the two $SL(2,Z)$'s of the torus
acting on $\rho$ and $\tau$ respectively.

Applying a T-duality: $R_1 \rightarrow {1 \over R_1}$ or equivalently 
$ \tau\leftrightarrow \rho$,
we obtain the mass spectrum of the open strings compactified on a 
noncommutative torus. Moreover, we observe that:  
\be
F^{-1}={n_1 \over n_2\rho_2}+ \cot \alpha.
\ee

The advantage of T-duality apart from providing the spectrum on noncommutative
torus, is to give its dependence on the integers $n_1$ and $n_2$ which now 
can be interpreted as the number of times the membrane is wrapped.

\section{M(atrix) Theory on $T^{\theta}$}

We have observed that membranes in string theory,
with mixed boundary condition, exhibit noncommuting space coordinates;
the amount of noncommutativity determined by the Kalb-Ramond field on the
membrane.

In the strong coupling limit this field is a pull back of the three form 
$C$ of the M-theory with non-zero component $C_{012}$ [7].
Therefore the M(atrix) description of this configuration must involve
non-vanishing $C_{-12}$ as M(atrix) model is purported to be M-theory in the 
infinite momentum frame [8].
It remains to compactify [9] the M(atrix) theory of [8] on the two directions
$X^1$ and $X^2$ i.e, to find operators $U_1$ and $U_2$ with the property
[9] 

\be 
U_i X_i U_i^{-1}=X_i + R_i. 
\ee
But now, the coordinates must also satisfy: 
\be
[X^1, X^2]=-2\pi i F.
\ee
It is the standard noncommutative torus as discussed in [2,3], with the 
$C^*$ algebra defining the geometry, generated by one pair of elements 
$U_1$ and $U_2$ as in eq.(1), with 
\be
\theta=\rho_2 \cot\alpha.
\ee

It is straightforward to see that the mass spectrum for the mixed membranes
is thus translated to the M(atrix) model context and reproduces the spectrum
of [2]. The $SL(2,Z)$ symmetries discussed in  section 3 
are therefore 
also the symmetries of the M(atrix) theory.
Specifically, after T-duality, our $SL(2,Z)$ acting on the $\rho$ modulus 
becomes the non-classical $SL(2,Z)_N$ of [2].

\end{document}